\documentclass{phb-proc4-auth}

\usepackage{graphicx}
\usepackage{amssymb}

\begin{document}
\begin{frontmatter}


\journal{SCES '04}


\title{Mott--Hubbard and Anderson Transitions in Dynamical Mean--Field Theory}

%
%
%
%
%
%

\author[kb]{Krzysztof Byczuk\corauthref{11}}
\author[wh]{Walter Hofstetter}
\author[dv]{Dieter Vollhardt}

%

\address[kb]{Institute of Theoretical Physics,
Warsaw University, ul. Ho\.za 69, PL-00-681 Warszawa, Poland}
\address[wh]{Condensed Matter Theory Group, Massachusetts Institute of Technology,
Cambridge, MA 02139, USA}
\address[dv]{Theoretical Physics III, Center for Electronic Correlations and Magnetism,
Institute for Physics,University of Augsburg, D-86135 Augsburg, Germany}

%
%
%
%


%
%
%
%

\corauth[11]{Corresponding Author: Institute of Theoretical Physics,
Warsaw University, ul. Ho\.za 69, PL-00-681 Warszawa, Poland. Phone: (48) 22
5532256, Fax: (48) 22  6219475, Email: byczuk@fuw.edu.pl}


\begin{abstract}
The Anderson--Hubbard Hamiltonian at half--filling is
investigated within dynamical mean--field theory at zero
temperature. The local density of states is calculated by taking
the geometric and arithmetic mean, respectively.
The non--magnetic ground state phase diagrams obtained within the different
averaging schemes are compared.

\end{abstract}

%
%

\begin{keyword}
Mott--Hubbard and Anderson Metal--Insulator Transitions,
  Dynamical Mean--Field Theory
\end{keyword}


\end{frontmatter}

%
%
%
%
%


The metal insulator transitions (MIT) due to electronic interactions
(Mott-Hubbard MIT) \cite{Mott90} and due to impurity scattering (Anderson
localization) \cite{Anderson58} are
subtle quantum mechanical phenomena which require non-perturbative
investigation schemes.
In this respect the dynamical mean-field theory (DMFT) \cite{georges96} is
very useful.
However, it cannot describe the physics of
Anderson localization if the local disorder is included by taking the
arithmetic average over the disorder \cite{ulmke95}.
Recently it was proposed to use the geometric average over the disorder
to include also the Anderson MIT \cite{Dobrosavljevic97,Dobrosavljevic03,Schubert03}.
In this contribution we compare
 the entire non-magnetic ground state phase diagrams
of correlated, disordered electrons at half filled lattice as obtained by
taking
the geometric  and the arithmetic averages over the disorder
within the DMFT \cite{byczuk04}.

We study the system described by a single--orbital Anderson--Hubbard model
\begin{equation}
H_{AH}=-t\sum_{\langle ij\rangle \sigma }a_{i\sigma }^{\dagger }a_{j\sigma
}+\sum_{i\sigma }\epsilon _{i}n_{i\sigma }+U\sum_{i}n_{i\uparrow
}n_{i\downarrow },  \label{1}
\end{equation}
where $t>0$ is the amplitude for hopping between nearest neighbors, $U$ is
the on--site repulsion, $n_{i\sigma }=a_{i\sigma }^{\dagger }a_{i\sigma }^{{%
\phantom{\dagger}}}$ is the local electron number operator,
$a_{i\sigma }$ ($a_{i\sigma }^{\dagger}$) is the annihilation
(creation) operator of an electron with spin $\sigma$, and the
local ionic energies $\epsilon _{i}$ are independent random
variables. We assume a continuous probability distribution for
$\epsilon _{i}$, i.e., $\mathcal{P}(\epsilon _{i})=\Theta (\Delta
/2-|\epsilon _{i}|)/\Delta ,$ with $\Theta $ as the step function.
Here,
 $\Delta $ is a measure of the disorder strength.

This model is solved within DMFT by mapping it \cite{georges96} onto an ensemble of
effective single--impurity Anderson Hamiltonians with different
$\epsilon _{i}$:
\begin{eqnarray}
H_{\mathrm{SIAM}} &=&\sum_{\sigma }(\epsilon _{i}-\mu )a_{i\sigma }^{\dagger
}a_{i\sigma }+Un_{i\uparrow }n_{i\downarrow }  \label{2} \\
&&+\sum_{\mathbf{k}\sigma }V_{\mathbf{k}}a_{i\sigma }^{\dagger }c_{\mathbf{k}%
\sigma }+V_{\mathbf{k}}^{\ast }c_{\mathbf{k}\sigma }^{\dagger }a_{i\sigma
}+\sum_{\mathbf{k}\sigma }\epsilon _{\mathbf{k}}c_{\mathbf{k}\sigma
}^{\dagger }c_{\mathbf{k}\sigma }.  \nonumber
\end{eqnarray}
Here $\mu = U/2$ is the chemical potential corresponding to a half-filled
band, and $V_{\mathbf{k}}$ and $\epsilon _{\mathbf{k}}$ are the
hybridization matrix element and the dispersion relation of the auxiliary
bath fermions $c_{\mathbf{k}\sigma }$, respectively.

For each ionic energy $\epsilon _{i}$ we calculate the local Green
function $G(\omega ,\epsilon _{i})$, from which we can obtain
either the geometrically averaged local density of states (LDOS)
$\rho _{\mathrm{geom}}(\omega )=\exp \left[ \langle \ln \rho
_{i}(\omega )\rangle \right]$ or the arithmetically averaged LDOS
$\rho _{\mathrm{arith}}(\omega )= \langle \rho _{i}(\omega
)\rangle  $, where $\rho _{i}(\omega )=-\mathrm{{Im}}G(\omega
,\epsilon _{i})/\pi $, and $\langle O_{i}\rangle =\int d\epsilon
_{i}\mathcal{P}(\epsilon _{i})O(\epsilon _{i})$ is the arithmetic
mean of $O_{i}$. The lattice Green function is given by the
corresponding Hilbert transform as $G(\omega )=\int d\omega
^{\prime }\rho _{\mathrm{\alpha}}(\omega ' )/(\omega -\omega
^{\prime })$, where the subscript $\mathrm{\alpha}$ stands for
either "$\mathrm{geom}$" or "$\mathrm{arith}$". The local
self--energy $\Sigma (\omega )$ is determined from the
$\mathbf{k}$-integrated Dyson equation $\Sigma(\omega )=\omega
-\eta (\omega )-1/G(\omega )$ where the hybridization function
$\eta (\omega )$ is defined as $\eta (\omega
)=\sum_{\mathbf{k}}|V_{\mathbf{k}}|^{2}/\left( \omega -\epsilon
_{\mathbf{k}}\right) $. The self--consistent DMFT equations are
closed through the Hilbert transform $ G(\omega )=\int d\epsilon
N_{0}(\epsilon )/\left[\omega -\epsilon -\Sigma (\omega )\right]
$, which relates the local Green function for a given lattice to
the self--energy; here $N_{0}(\epsilon )$ is the non--interacting
DOS.

The Anderson--Hubbard model (\ref{1}) is solved for
a semi-elliptic DOS, $N_{0}(\epsilon )=4\sqrt{1-4\epsilon ^{2}}/\pi$.
Then $\eta (\omega )=G(\omega )/16$.
 The DMFT equations are solved
at zero temperature by the numerical renormalization group technique
\cite{NRG}.
For numerical integrations  we use discrete values of $\epsilon_i$
selected according to the Gauss-Legendre algorithm.
The number of $\epsilon_i$ levels depends on $\Delta$ 
and is adjusted to obtain  smooth spectral functions \cite{ulmke95}.

 \begin{figure}
     \centering
     \includegraphics[width=0.4\textwidth]{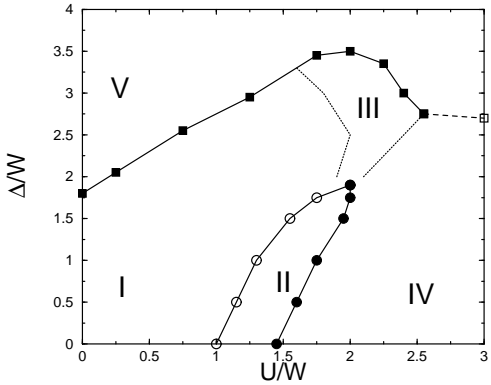}
     \caption{Non--magnetic ground state phase diagram of the Anderson--Hubbard model at
half-filling as calculated by DMFT with the geometrically averaged  local density of states.
I -- correlated, disordered metal, II -- coexistence regime, III -- crossover
regime, IV -- Mott insulator, V -- Anderson insulator. At the dashed line all
states within the Hubbard subbands become localized. $W$ is a bare energy
band-width.}
 \end{figure}

 \begin{figure}
     \centering
     \includegraphics[width=0.4\textwidth]{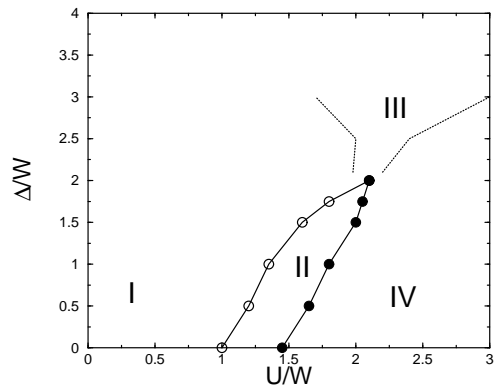}
     \caption{Non--magnetic ground state phase diagram of the Anderson--Hubbard model at
half-filling as calculated by DMFT with the arithmetically averaged local
density of states.
Labels I -- IV are the same as in Fig. 1.}
 \end{figure}

The main results are presented in Fig.~1 and 2, where the
non--magnetic ground state phase diagrams of the Anderson--Hubbard
model (1) are plotted. At weak disorder both averaging schemes are
seen to lead to very similar results for the Mott-Hubbard
transition from the paramagnetic correlated, disordered  metal (I)
to the paramagnetic, disordered Mott insulator (IV). For
$\Delta\lesssim 2$ we find hysteresis and a coexistence regime
(II). Above $\Delta\approx 2$ there is a crossover regime (III).
At strong disorder the DMFT with geometrically averaged LDOS
describes an Anderson transition between a correlated, disordered
metallic and insulating phase, respectively (V) \cite{comment}. 
Since neither the
paramagnetic Mott nor the Anderson insulator is characterized by a
broken symmetry they are continously connected, i.e., the phases
IV and V in Fig.~1 are connected by a continuous line which does
not cross a metallic phase. Obviously, the Anderson transition is
missed within the DMFT supplied by the arithmetic averaging.

In summary, although the geometric averaging procedure does not
capture all aspects of Anderson localization
\cite{Dobrosavljevic03,alvermann04} it provides valuable new
insights into MITs in correlated and disordered electron systems,
which are not obtained by taking the arithmetic disorder average
\cite{ulmke95}.

This work was supported in part by the
Sonderforschungsbereich 484 of the Deutsche Forschungsgemeinschaft (DFG).
Financial support of KB through KBN-2 P03B 08 224, and of WH through the DFG
and a Pappalardo Fellowship is gratefully acknowledged.

%
%
%
%

%
%
%
%



\begin{thebibliography}{00}


\bibitem{Mott90} N.~F.~Mott, Proc. Phys. Soc. A {\bf 62}, 416 (1949);
{\em Metal--Insulator Transitions}, 2nd edn. (Taylor and Francis,
London 1990).

\bibitem{Anderson58} P.~W.~Anderson, Phys. Rev. {\bf 109}, 1492 (1958).

\bibitem{georges96} A.~Georges {\it et al.},
Rev.\ Mod.\ Phys.\ {\bf 68}, 13 (1996).


\bibitem{ulmke95} M.~Ulmke, V.~Jani\v{s}, and D.~Vollhardt, Phys. Rev. B {\bf 51}, 10411 (1995).



\bibitem{Dobrosavljevic97} V.~Dobrosavljevi\'c and G.~Kotliar, Phys. Rev. Lett. {\bf 78}, 3943 (1997).

\bibitem{Dobrosavljevic03} V.~Dobrosavljevi\'c, A.~A.~Pastor, and B.~K.~Nikoli\'c,
Europhys. Lett. {\bf 62}, 76 (2003).

\bibitem{Schubert03} G.~Schubert, A.~Wei{\ss}e, and H.~Fehske, cond-mat/0309015.

\bibitem{byczuk04} K.~Byczuk, W.~Hofstetter, and D.~Vollhardt,
  cond-mat/0403765, Phys. Rev. Lett. {\bf 94}, (2005) - in press.

\bibitem{NRG}
 R.~Bulla, Phys. Rev. Lett. {\bf 83}, 136 (1999).

\bibitem{comment} Currently we try to improve the results when $U\to 0$ using
  different route to obtain the self-energy. This should reduce
  effects due to a logarithmic broadening within NRG scheme \cite{NRG}. 

\bibitem{alvermann04} A.~Alvermann, {\it at al.}, cond-mat/0406051.



\end{thebibliography}
\end{document}